\documentclass[aps,showpacs]{revtex4}

\usepackage{amsmath,amsfonts}
\usepackage[dvips]{graphicx}

\begin{document}

\title{Complementary constraints on non-standard cosmological models \\ 
from CMB and BBN}
\author{Adam Krawiec}
\email{uukrawie@cyf-kr.edu.pl}
\affiliation{Institute of Public Affairs, Jagiellonian University,
Rynek G{\l}{\'o}wny, 31-042 Krak{\'o}w, Poland}
\author{Marek Szyd{\l}owski}
\email{uoszydlo@cyf-kr.edu.pl}
\affiliation{Complex Systems Research Centre, Jagiellonian University,
Reymonta 4, 30-059 Krak{\'o}w, Poland}
\author{W{\l}odzimierz God{\l}owski}
\email{godlows@oa.uj.edu.pl}
\affiliation{Astronomical Observatory, Jagiellonian University,
Orla 171, 30-244 Krak{\'o}w, Poland}

\begin{abstract}
We study metric-affine gravity (MAG) inspired cosmological models. Those 
models were statistically estimated using the SNIa data. We also use the 
cosmic microwave background observations and 
the big-bang nucleosynthesis analysis to constrain the density parameter 
$\Omega_{\psi,0}$ which is related to the non-Riemannian structure of the 
underlying spacetime. We argue that while the models are statistically 
admissible from the SNIa analysis, complementary stricter limits obtained 
from the CMB and BBN indicate that the models with density parameters 
with a $a^{-6}$ scaling behaviour are virtually ruled out. If we assume the 
validity of the particular MAG based cosmological model throughout all 
stages of the Universe, the parameter estimates from the CMB and BBN put 
a stronger limit, in comparison to the SNIa data, on the presence of 
non-Riemannian structures at low redshifts.
\end{abstract}


\maketitle

\section{Introduction}

Astronomical observations brought important changes in modern cosmology
\cite{Dodelson:2003}. While the type Ia supernovae (SNIa) data are most often
employed, the cosmic microwave background (CMB) observations and big-bang 
nucleosynthesis (BBN) analysis can also be used. They allow the exotic 
physics of cosmological models to be checked against the observational
data \cite{Kamionkowski:2002}. There is an increasing effort to develop some 
cosmological and astrophysical tools to search for new physics beyond the 
standard model.

The metric-affine gravity (MAG) cosmological model with the Robertson-Walker
symmetry was investigated by three different groups. First, it was
considered the model with triplet ansatz in vacuum \cite{Obukhov:1997}.
Second, it was considered the dilational hyperfluid model \cite{Babourova:2003}.
Third, it was shown that on the level of the field equations the special
case of the MAG model is equivalent to a model in the Weyl-Cartan spacetime
if we choose a model parameter in the special form ($a_6 = - a_4$)
\cite{Puetzfeld:2004}. Moreover after redefinition of some variables
the second and third approach gives the same set of dynamical equations.
The analysis of constraints on parameters in the MAG model can be addressed 
in all three approach, but we adopt the last one proposed by Puetzfeld and 
Chen \cite{Puetzfeld:2004}.

Note that in the model with dust matter on the brane, apart from dark radiation 
which scales like $a^{-4}$, there is a correction of the type $a^{-6}$ to the 
Einstein equations on the brane which arise from the influence of a bulk 
geometry \cite{Randall:1999,Randall:1999b,Godlowski:2004}. The term scaling 
like $a^{-6}$ also appears in the Friedmann-Robertson-Walker model with 
spinning fluid \cite{Szydlowski:2004c}. It is possible to establish formally 
the one to one correspondence between the MAG model and either the 
Randall-Sundrum brane model when positive values of the non-Riemannian 
contribution to effective energy is admitted or the spinning fluid filled 
cosmology when this contribution is negative. However, if one takes the pure 
Randall-Sundrum type model then there is a constraint on the brane tension 
parameter coming from the theory itself. The brane tension parameter is not 
less than about ($100$ GeV)${}^{4}$. The MAG model is free from such a 
theoretical constraint. 

In our further discussion we examine the flat models which is motivated by 
the CMB WMAP observations \cite{Kamionkowski:1994} and consider the following 
formula for the Friedmann first integral
\begin{equation}
\label{eq:1}
\frac{H^{2}}{H_{0}^{2}} = \Omega_{{\rm m},0} a^{-3} + \Omega_{\Lambda,0} +
\Omega_{{\rm r},0} a^{-4} + \Omega_{\psi,0} a^{-6}
\end{equation}
where $H= d\ln a/d t$ is the Hubble function, $t$ is the cosmological time, 
$a=a(t)$ is the scale factor, $\Omega_{{\rm m},0}$, $\Omega_{\Lambda,0}$ 
and $\Omega_{\psi,0}$ are the density parameters for dust matter, the 
cosmological constant and fictitious fluid which mimics ``non-Riemannian 
effects'', respectively. Their values in the present epoch are marked by the 
index ``0''. All density parameters satisfy the constraint condition 
\begin{equation}
\label{eq:2}
\Omega_{{\rm m},0} + \Omega_{\Lambda,0} 
+ \Omega_{{\rm r},0} + \Omega_{\psi,0} = 1.
\end{equation}
The density parameter for the fictitious fluid is defined 
as \cite{Puetzfeld:2004}
\begin{equation}
\label{eq:3}
\Omega_{\psi} = \frac{\upsilon}{H^2} \frac{\psi^2}{a^6}
\end{equation}
where
\[
\upsilon = \frac{\kappa^2}{144a_{0}} \left( 1 - \frac{3a_0}{b_4} \right).
\]
The sign of the parameter $\upsilon$ is undetermined and it can assume both 
positive and negative values.

\section{Constraint from the SNIa}

Let us start from the reestimation of the models parameters by using
the latest sample of SNIa data \cite{Riess:2004}. The motivation to
study the SN constraint is to find the best estimation available from
the latest data which gives the narrowest constraint for this method.
In the next sections we compare it with constraints obtained from other
methods.

Riess et al.'s sample contains 157 type Ia supernovae \cite{Riess:2004}.
We consider the flat model with and without priors on the $\Omega_{\psi,0}$
and $\Omega_{{\rm m},0}$. We assume that the former can be of any value or 
only nonnegative, and the latter is nonnegative or equal $0.3$ 
\cite{Peebles:2003}. We estimate the best fits of the model parameters 
(Table~\ref{tab:1}). Additionally we find the maximum likelihood estimates 
with the errors at $2\sigma$ level (Table~\ref{tab:2}).

\begin{table}
\caption{Best fit estimation of the model parameters from the SNIa data}
\label{tab:1}
\begin{tabular}{@{}l|cccccc}
\hline
priors & $\Omega_{{\rm m},0}$ & $\Omega_{{\rm r},0}$ & $\Omega_{\psi,0}$ &
$\Omega_{\Lambda,0}$ & $\mathcal{M}$ & $\chi^{2}$ \\ \hline
$\Omega_{\psi,0} \ge 0$; $\Omega_{{\rm m},0} \ge 0$ &
$0$ & $0.14$ & $0.012$ & $0.848$ & $15.945$ & $175.75$ \\
$\Omega_{\psi,0} \ge 0$; $\Omega_{{\rm m},0} = 0.3$ &
--- & $0$ & $0.005$ & $0.695$ & $15.965$ & $177.30$ \\
$\Omega_{{\rm m},0} \ge 0$ &
$0$ & $0.14$ & $0.012$ & $0.848$ & $15.945$ & $175.75$ \\
$\Omega_{{\rm m},0} = 0.3$ &
--- & $0$ & $0.005$ & $0.695$ & $15.965$ & $177.30$ \\
$\Omega_{{\rm m},0} \ge 0$; $\Omega_{{\rm r},0}=0.0001$ &
$0.16$ & --- & $0.029$ & $0.811$ & $15.945$ & $175.97$ \\
$\Omega_{{\rm m},0} = 0.3$; $\Omega_{{\rm r},0}=0.0001$ &
--- & --- & $0.005$ & $0.695$ & $15.965$ & $177.30$ \\ \hline
\end{tabular}
\end{table}

\begin{table}
\label{tab:2}
\caption{Maximum likelihood estimation of the model parameters with $2\sigma$
errors from the SNIa data}
\begin{tabular}{@{}l|cccccc}
\hline
priors & $\Omega_{{\rm m},0}$ & $\Omega_{{\rm r},0}$ & $\Omega_{\psi,0}$ &
$\Omega_{\Lambda,0}$  \\ \hline
$\Omega_{\psi,0} \ge 0$; $\Omega_{{\rm m},0} \ge 0$ &
$0_{-0.00}^{+0.24}$ & $0_{-0.00}^{+0.15}$ & $0.009_{-0.009}^{+0.041}$ &
$0.820_{-0.120}^{+0.100}$ \\
$\Omega_{\psi,0} \ge 0$; $\Omega_{{\rm m},0} = 0.3$ & --- & 
$0_{-0.00}^{+0.04}$ & $0_{-0.000}^{+0.017}$ & $0.680_{-0.040}^{+0.020}$ \\
$\Omega_{{\rm m},0} \ge 0$ &
$0_{-0.00}^{+0.27}$ & $0_{-0.00}^{+0.21}$ & $0.009_{-0.036}^{+0.042}$ &
$0.800_{-0.140}^{+0.110}$ \\
$\Omega_{{\rm m},0} = 0.3$ & --- & $0_{-0.00}^{+0.09}$ & 
$\-0.002_{-0.022}^{+0.017}$ & $0.680_{-0.060}^{+0.020}$ \\
$\Omega_{{\rm m},0} \ge 0$; $\Omega_{{\rm r},0}=0.0001$ &
$0.14_{-0.14}^{+0.28}$ & --- & $0.028_{-0.035}^{+0.038}$ & 
$0.810_{-0.150}^{+0.130}$ \\
$\Omega_{{\rm m},0} = 0.3$; $\Omega_{{\rm r},0}=0.0001$ &
--- & --- & $0.005_{-0.013}^{+0.013}$ & $0.695_{-0.015}^{+0.015}$ \\ \hline
\end{tabular}
\end{table}

We find that that the estimates of the parameter $\Omega_{\psi,0}$ are
very close to zero although positive apart of one case when it is zero.
We can conclude that the estimate of this parameter is order of magnitude of
$0.01$.

\begin{figure}
\begin{center}
\includegraphics[width=0.48\textwidth]{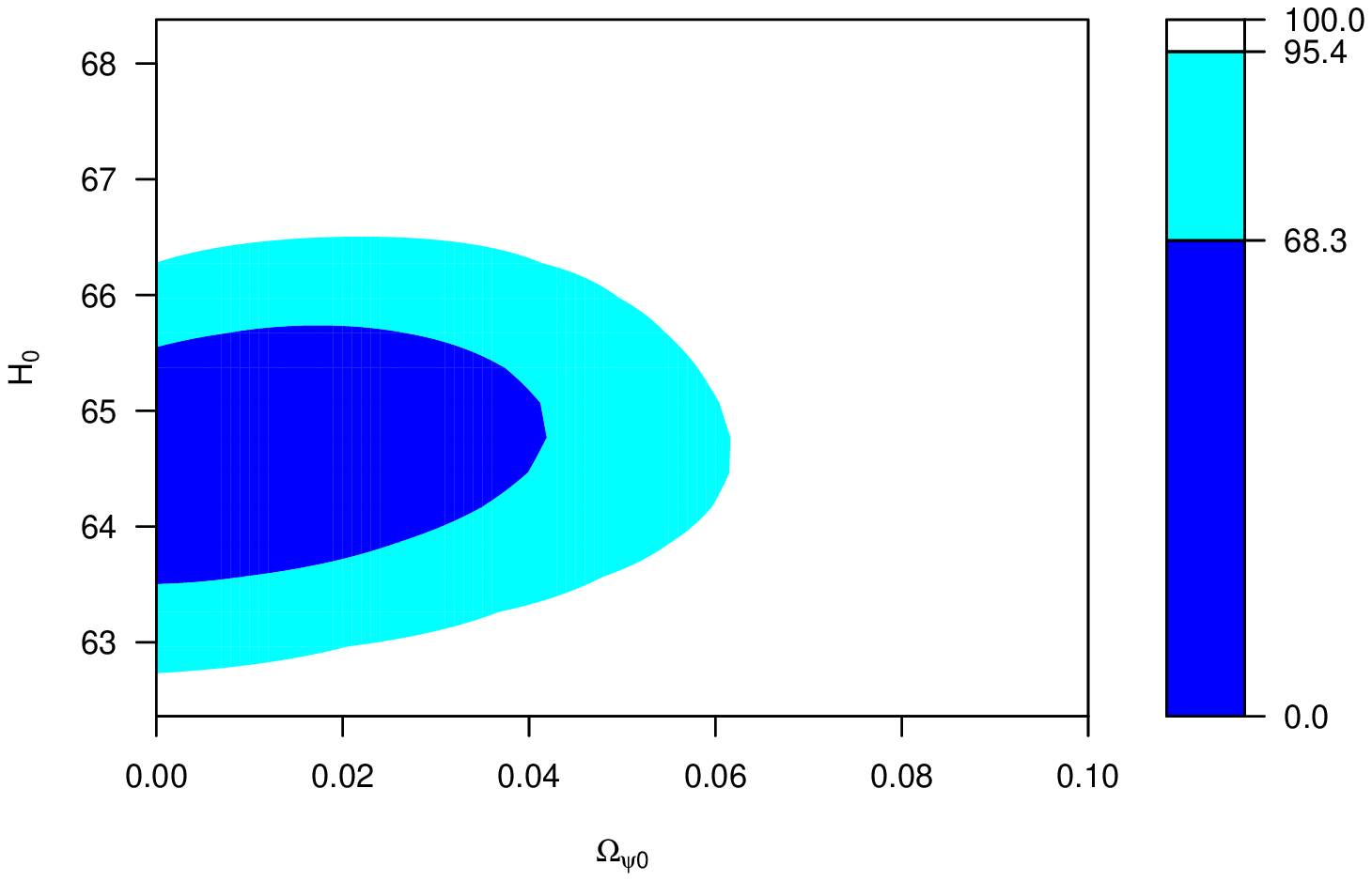}
\includegraphics[width=0.48\textwidth]{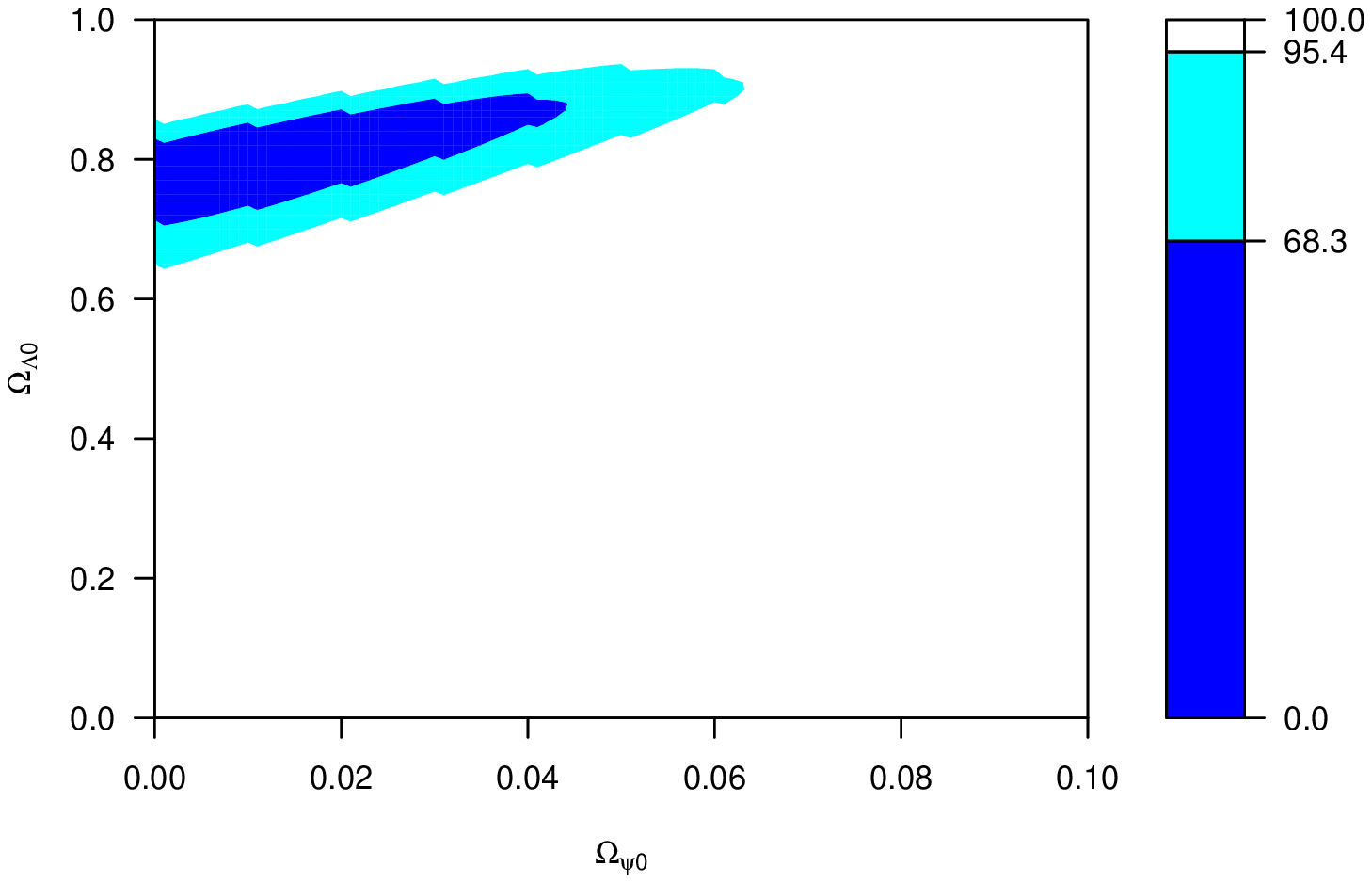} \\
\includegraphics[width=0.48\textwidth]{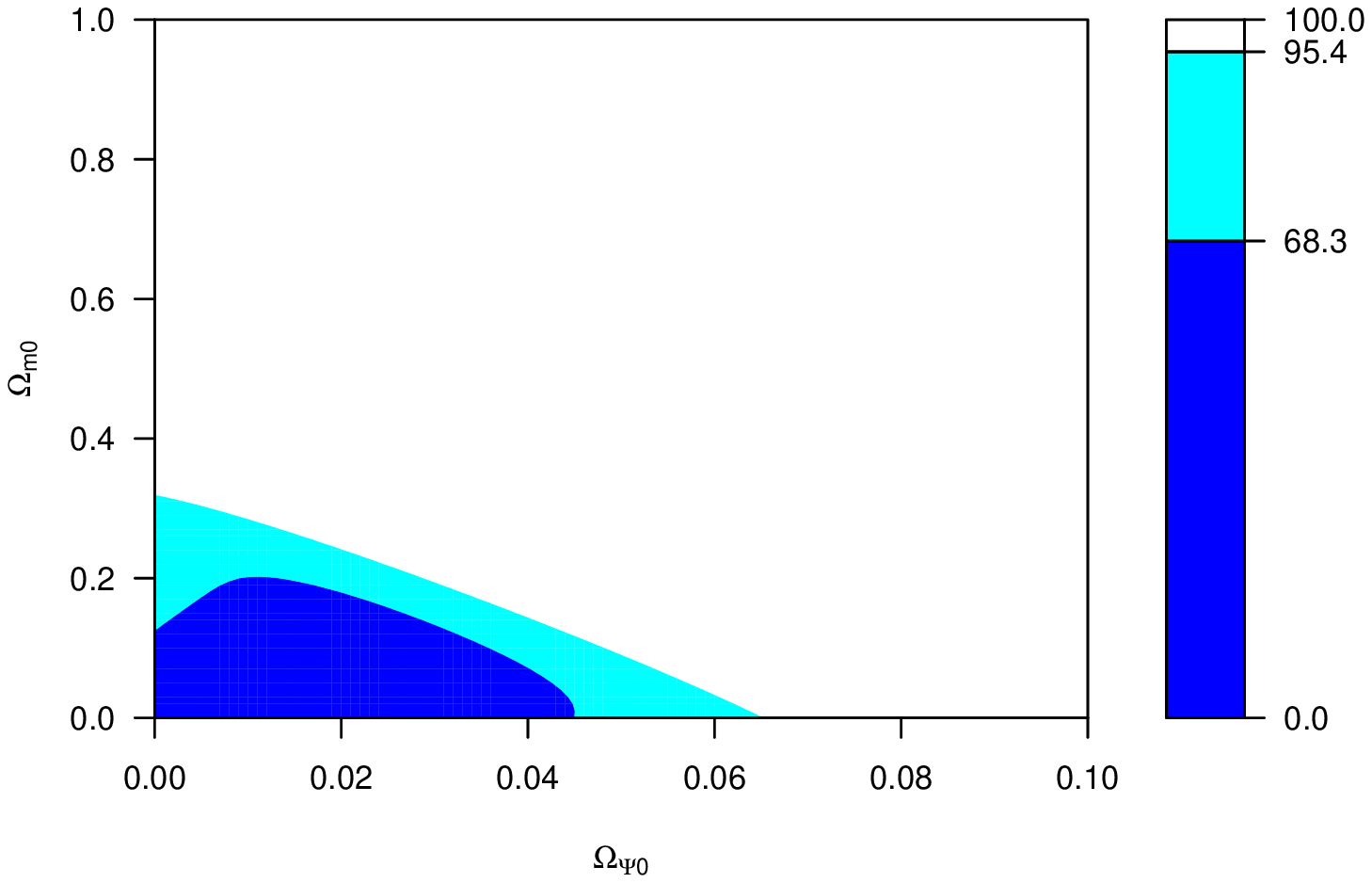}
\includegraphics[width=0.48\textwidth]{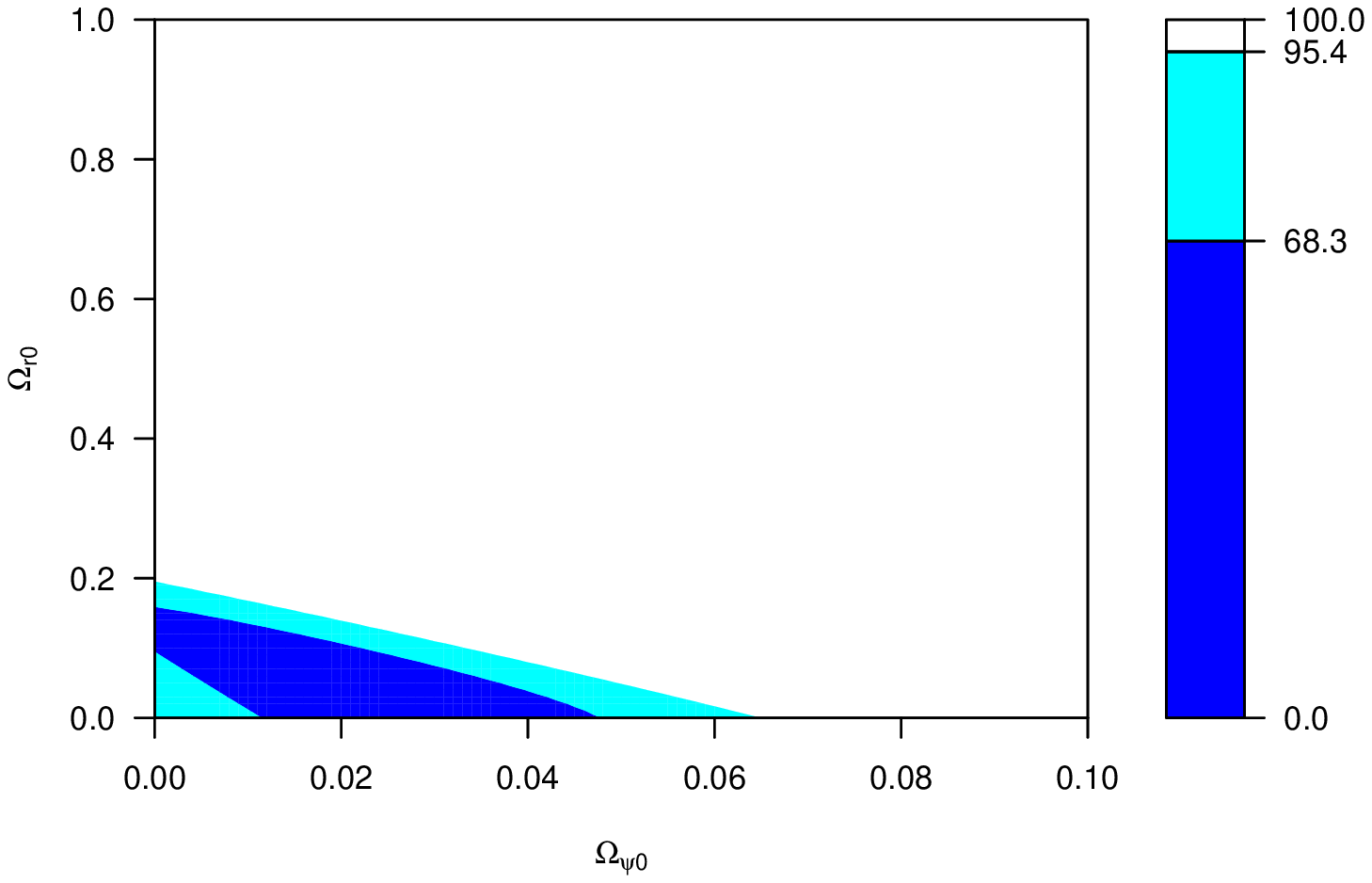}
\end{center}
\caption{The contours with $1\sigma$ and $2\sigma$ confidence levels 
for $\Omega_{\psi,0}$ versus $H_{0}$, $\Omega_{\Lambda,0}$, 
$\Omega_{{\rm m},0}$, and $\Omega_{{\rm r},0}$ from the SNIa data.}
\label{fig:1}
\end{figure}

To illustrate the results of the maximum likelihood analysis of the model
we draw the levels of confidence on Figure~\ref{fig:1}.

The MAG model fits well to SNIa data. We consider the model with any value of 
$\Omega_{{\rm r},0}$ we obtain the value of $\Omega_{{\rm m},0}$ equal zero 
as best fit and maximum likelihood estimator, while fixing the small amount 
radiation ($\Omega_{{\rm r},0}$ \cite{Vishwakarma:2003}) gives the low 
density matter universe. The estimation of the Hubble constant gives the 
value close to $65$ km/s MPc.

\section{CMB peaks in the MAG model}

The hotter and colder spots in the CMB can interpreted as acoustic oscillation
in the primeval plasma during the last scattering. Peaks in the power spectrum
correspond to maximum density of the wave. In the Legendre multipole space
these peaks correspond to the angle subtended by the sound horizon at the
last scattering. Further peaks answer to higher harmonics of the principal
oscillations.

It is very interesting that locations of these peaks are very sensitive to
the variations in the model parameters. Therefore, it can be used as another
way to constrain the parameters of cosmological models.

The acoustic scale $\ell_{A}$ which puts the locations of the peaks is defined
as
\begin{equation}
\label{eq:4}
\ell_{A} = \pi \frac{\int_{0}^{z_{\rm dec}} \frac{d z'}{H(z')}}
{\int_{z_{\rm dec}}^{\infty} c_{s} \frac{d z'}{H(z')}}
\end{equation}
where
\begin{equation}
\label{eq:5}
H(z) = H_{0} \sqrt{\Omega_{{\rm m},0}(1+z)^3 + \Omega_{{\rm r},0}(1+z)^4+
\Omega_{\psi,0}(1+z)^6+\Omega_{\Lambda,0}}
\end{equation}
and $c_{\rm s}$ is the speed of sound in the plasma given by
\begin{equation}
\label{eq:6}
c_{\rm s}^{2} \equiv \frac{d p_{\rm eff}}{d \rho_{\rm eff}} =
\frac{\frac{4}{3} \Omega_{\gamma,0}(1+z) + 6 \Omega_{\psi,0}(1+z)^{3}}
{3 \Omega_{{\rm b},0} + 4 \Omega_{\gamma,0}(1+z) + 6\Omega_{\psi,0}(1+z)^{3}}.
\end{equation}
Knowing the acoustic scale we can determine the location of $m$-th peak
\begin{equation}
\label{eq:7}
\ell_{m} \sim \ell_{A}(m- \phi_{m})
\end{equation}
where $\phi_{m}$ is the phase shift caused by the plasma driving effect.
Assuming that $\Omega_{{\rm m},0}=0.3$, on the surface of last scattering
$z_{\rm dec}$ it is given by
\begin{equation}
\label{eq:8}
\phi_{m} \sim 0.267 \left[ \frac{r(z_{\rm dec})}{0.3} \right]^{0.1} =
0.267 \left[ \frac{1}{0.3} \frac{\rho_{\rm r}(z_{\rm dec})}
{\rho_{\rm m}(z_{\rm dec})} \right]^{0.1} =
0.267 \left[ \frac{1}{0.3} \frac{\Omega_{{\rm r},0}(1+z_{\rm dec})}
{0.3} \right]^{0.1}
\end{equation}
where $\Omega_{{\rm b},0}h^{2}=0.02$, $r(z_{\rm dec}) \equiv 
\rho_{\rm r}(z_{\rm dec})/\rho_{\rm m}(z_{\rm dec}) =
\Omega_{{\rm r},0}(1+z_{\rm dec})/\Omega_{{\rm m},0}$ is the ratio 
of radiation to matter densities at the surface of last scattering. 

The CMB temperature angular power spectrum provides the locations of the first 
two peaks \cite{Spergel:2003,Page:2003} and the BOOMERanG measurements give 
the location of the third peak \cite{deBernardis:2002}. They values with
uncertainties on the level 1$\sigma$ are the following
\[
\ell_{1} = 220.1_{-0.8}^{+0.8}, \qquad
\ell_{2} = 546_{-10}^{+10}, \qquad
\ell_{3} = 845_{-25}^{+12}.
\]
Using the WMAP data only, Spergel et al. \cite{Spergel:2003} obtained that 
the Hubble constant $H_{0}=72$ km/s MPc (or the parameter $h=0.72$), the 
baryonic matter density $\Omega_{{\rm b},0} = 0.024h^{-2}$, and the matter 
density $\Omega_{{\rm m},0} = 0.14h^{-2}$ which give a good agreement with 
the observation of position of the first peak. 

To find whether cosmological models give these observational locations of 
peaks we fix some model parameters. Let the baryonic matter density 
$\Omega_{{\rm b},0}=0.05$, the spectral index for initial density 
perturbations $n=1$, and the radiation density parameter 
\cite{Vishwakarma:2003}
\begin{equation}
\label{eq:9}
\Omega_{{\rm r},0} = \Omega_{\gamma,0} + \Omega_{\nu,0} 
= 2.48 h^{-2} \times 10^{-5} + 1.7 h^{-2} \times 10^{-5} 
= 4.18 h^{-2} \times 10^{-5}
\end{equation}
which is a sum of the photon $\Omega_{\gamma,0}$ and neutrino $\Omega_{\nu,0}$ 
densities. 

Assuming $\Omega_{{\rm m},0} = 0.3$ and $h=0.72$ we obtain for the standard 
$\Lambda$CDM cosmological model the following positions of peaks 
\[
\ell_{1} = 220, \qquad
\ell_{2} = 521, \qquad
\ell_{3} = 821
\]
with the phase shift $\phi_{m}$ given by (\ref{eq:8}).

From the SNIa data analysis it was found that the Hubble constant 
has lower value. Assuming that $H_{0}=65$ km/s MPc (or $h=0.65$), we have 
$\Omega_{{\rm r},0} = 9.89 \times 10^{-5}$ from equation~(\ref{eq:9}). In 
further calculation we take $\Omega_{{\rm r},0} = 0.0001$. 
If we consider the standard $\Lambda$CDM model, with $\Omega_{{\rm m},0}=0.3$, 
$\Omega_{{\rm b},0}=0.05$, the spectral index for initial density 
perturbations $n=1$, and $h=0.65$, where sound can propagate in baryonic matter 
and photons we obtain the following locations of first three peaks
\[
\ell_{1} = 225, \qquad
\ell_{2} = 535, \qquad
\ell_{3} = 847.
\]
We find some discrepancy between the observational and theoretical results in
this case. Now it is interesting to check whether the presence of the 
fictitious fluid $\Omega_{\psi,0}$ change the locations of the peaks.

The properties of the fictitious fluid $\Omega_{\psi,0}$ are unknown. In 
particular, we do not know whether the sound can or cannot propagate in this 
fluid. But we assume that sound can propagate in it as well as in baryonic 
matter and photons. We consider both values of the Hubble constant and assume 
that $h=0.65$ or $h=0.72$. The results of calculations of peak locations 
and the values of the parameter $\Omega_{\psi,0}$ are presented in 
Table~\ref{tab:3}.

\begin{table}
\caption{Values of 
$\Omega_{\psi,0}$ and location of first three peaks}
\label{tab:3}
\begin{tabular}{@{}c|cccc}
\hline
Hubble constant & $\Omega_{\psi,0}$ & $\ell_{1}$ & $\ell_{2}$ & $\ell_{3}$ \\ \hline
$H_{0} = 65$ km/s MPc 
& $3 \times 10^{-11}$ & $220$ & $522$ & $825$ \\
& $7 \times 10^{-14}$ & $220$ & $523$ & $826$ \\
& $-1.4 \times 10^{-10}$ & $223$ & $530$ & $847$ \\
$H_{0} = 72$ km/s MPc
& $3.7 \times 10^{-11}$ & $220$ & $522$ & $823$ \\
& $0$ & $220$ & $521$ & $821$ \\
& $-1.3 \times 10^{-10}$ & $224$ & $530$ & $847$ \\
\hline
\end{tabular}
\end{table}

\begin{figure}
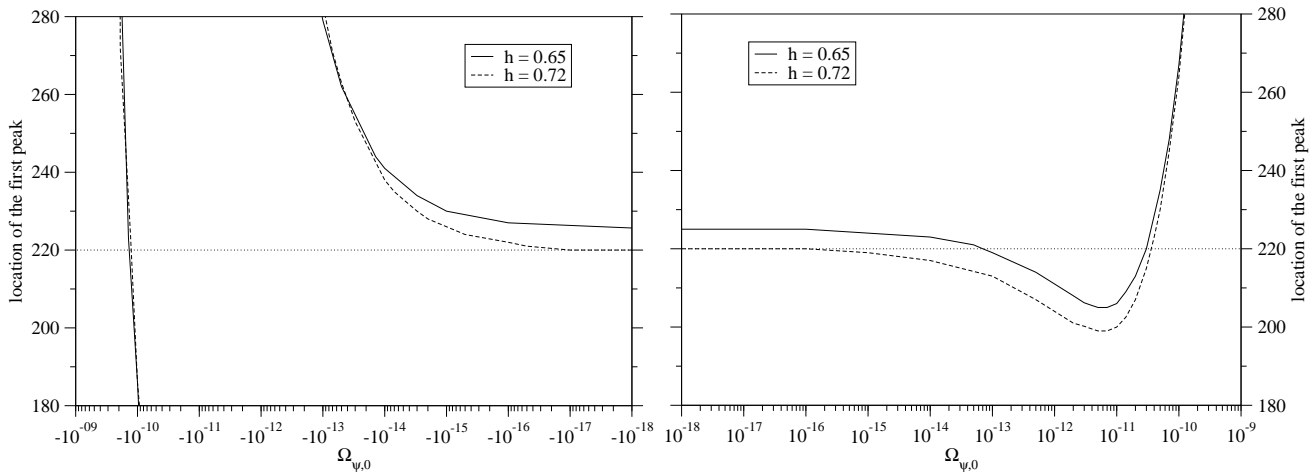

\includegraphics[width=0.48\textwidth]{fig2b.eps}
\includegraphics[width=0.48\textwidth]{fig2a.eps}
\caption{The location of the first peak in function of
$\Omega_{\psi,0}$.}
\label{fig:2}
\end{figure}

If we choose the $H_0=65$ km/s MPc then we obtain the agreement with the 
observation of the location of the first peak for three non-zero values of 
the parameter $\Omega_{\psi,0}$. As it is shown on Figure~\ref{fig:2} there 
are two positive and one negative values of this parameter for which the MAG 
model is admissible. 

All these distinguished values of $\Omega_{\psi,0}$ are in agreement with the 
result obtained from SNIa because the $2\sigma$ confidence interval for 
this parameter obtained from the SNIa data contains these three points. 
While the SNIa estimations give the possibility that $\Omega_{\psi,0}$ is 
equal zero, the CMB calculations seem to exclude this 
case because the zero value of $\Omega_{\psi,0}$ requires the first peak 
location at $225$.

If we choose the $H_0=72$ km/s MPc than one of positive values of 
$\Omega_{\psi,0}$ move to zero, while the second one move a little to 
the right. 

We also calculated the age of the Universe in the MAG model. We find that the 
difference in the age of the Universe is smaller than 1 mln years for all 
three values of $\Omega_{\psi,0}$.  Assuming that $\Omega_{{\rm m},0}=0.3$ 
the age of the Universe is $14.496$ Gyr for $H_{0} = 65$ km/s MPc, and 
$13.088$ Gyr for $H_{0} = 72$ km/s MPc. The globular cluster analysis 
indicated that the age of the Universe is $13.4$ Gyr \cite{Chaboyer:2003}.

\section{Constraint from the BBN}

It is well known that the big-bang nucleosynthesis (BBN) is the very well 
tested area of cosmology and does not allow for any significant deviation 
from the standard expansion law apart from very early times (i.e., before 
the onset of BBN). The prediction of standard BBN is in well agreement 
with observations of abundance of light elements. Therefore, all nonstandard 
terms added to the Friedmann equation should give only negligible small 
modifications during the BBN epoch to have the nucleosynthesis process 
unchanged. 

In our opinion the consistency with BBN is a crucial issue in the MAG models 
where the nonstandard term $a^{-6}$ in the Friedmann equation is added (see 
also discussion in \cite{Puetzfeld:2003}). This 
additional term scales like $(1+z)^6$. It is clear that such a term has either 
accelerated ($\Omega_{\psi,0}>0$) or decelerated ($\Omega_{\psi,0}<0$) impact 
on the Universe expansion. Going backwards in time this term would become 
dominant at some redshift. If it would happen before the BBN epoch, the 
radiation domination would never occur and the all BBN predictions would be 
lost. 

The domination of the fictitious fluid $\Omega_{\psi}$ should end before the 
BBN epoch starts otherwise the nucleosynthesis process would be dramatically 
modified. If we assume that the BBN result are preserved in the MAG models we 
obtain another constraint on the amount of $\Omega_{\psi,0}$. Let us assume 
that the model modification is negligible small during the BBN epoch and 
the nucleosynthesis process is unchanged. It means that the contribution of 
the MAG term $\Omega_{\psi,0}$ cannot dominate over the radiation term 
$\Omega_{{\rm r},0} \approx 10^{-4}$ before the beginning of BBN 
($z \simeq 10^{8}$) 
\[
\Omega_{\psi,0}(1+z)^{6} < \Omega_{{\rm r},0}(1+z)^{4} 
\qquad \Longrightarrow \qquad |\Omega_{\psi,0}| < 10^{-20}.
\]

The values of $\Omega_{\psi,0}\propto 10^{-2}$ obtained as best fits in the 
SNIa data analysis as well as the smallest nonzero value of $\Omega_{\psi,0}= 
7\times 10^{-14}$ calculated in the CMB analysis are unrealistic in the light
of the above result. If we take into consideration the maximum likelihood 
analysis of SNIa data we have the possibility that the value of 
$\Omega_{\psi,0}$ is lower than $|10^{-20}|$ in the $2\sigma$ confidence 
interval. In the case of the CMB analysis only the value of the Hubble 
constant close to $72$ km/s MPc gives the zero or close to zero value of 
$\Omega_{\psi,0}$.

\section{Conclusion}

The paper discusses observational constraint on ``energy contributions'' 
arising in certain cosmological models based on MAG. In particular it is focused on the nonstandard 
term $a^{-6}$. We test this model against the SNIa data, the location of the 
peaks of the CMB power spectrum, and constraints from the BBN. 

The MAG model fits well to SNIa data and the estimations give the amount of 
fluid $\Omega_{\psi,0}$ to be order of magnitude $0.01$, and the Hubble 
constant is close to $65$ km/s MPc. Let us note that these results are 
compatible with constraints from FRIIb radio galaxies and X-ray gas mass 
fractions \cite{Puetzfeld:2005}.

The CMB analysis gives that the Hubble constant is $72$ km/s MPc which gives 
the too low age of the universe in comparison with the age of globular 
clusters. Taking lower value of the Hubble constant obtained from SNIa 
estimation resolves the problem of the age. However, the location of the first 
peak shifts to the right and is in conflict with the observed location. The 
introducing of the non-Riemannian structure of the underlying spacetime moves 
the location of the first peak back and this MAG model agrees with the CMB 
observations. The analysis of the CMB in this model cannot distinguish the 
character of the fictitious fluid and we do not know whether the parameter 
$\Omega_{\psi,0}$ is positive or negative. 

The absolute values of $\Omega_{\psi,0}$ obtained in the MAG model from the 
CMB analysis with $h=0.65$ seems to be too large in comparison to the limit 
obtained from the BBN analysis. Using the BBN analysis we pointed out that 
the MAG part of the energy density to its present density parameter is of 
order $10^{-20}$. The limit of this order leads to the conclusion that the 
MAG model is virtually ruled out. However, we must remember that we insist 
that the MAG model does not change the physics during and after the BBN epoch. 
In this context, the merit of the SNIa analysis is its independency from 
any assumption on physical processes in the early Universe.

\acknowledgments
M. Szydlowski acknowledges the support by KBN grant no. 1 P03D 003 26.


\begin{thebibliography}{19}
\expandafter\ifx\csname natexlab\endcsname\relax\def\natexlab#1{#1}\fi
\expandafter\ifx\csname bibnamefont\endcsname\relax
  \def\bibnamefont#1{#1}\fi
\expandafter\ifx\csname bibfnamefont\endcsname\relax
  \def\bibfnamefont#1{#1}\fi
\expandafter\ifx\csname citenamefont\endcsname\relax
  \def\citenamefont#1{#1}\fi
\expandafter\ifx\csname url\endcsname\relax
  \def\url#1{\texttt{#1}}\fi
\expandafter\ifx\csname urlprefix\endcsname\relax\def\urlprefix{URL }\fi
\providecommand{\bibinfo}[2]{#2}
\providecommand{\eprint}[2][]{\url{#2}}

\bibitem[{\citenamefont{Dodelson}(2003)}]{Dodelson:2003}
\bibinfo{author}{\bibfnamefont{S.}~\bibnamefont{Dodelson}},
  \emph{\bibinfo{title}{Modern Cosmology}} (\bibinfo{publisher}{Academic
  Press}, \bibinfo{address}{San Diego}, \bibinfo{year}{2003}).

\bibitem[{\citenamefont{Kamionkowski}(2002)}]{Kamionkowski:2002}
\bibinfo{author}{\bibfnamefont{M.}~\bibnamefont{Kamionkowski}},
  \bibinfo{journal}{ECONF} \textbf{\bibinfo{volume}{C020805}},
  \bibinfo{pages}{TF04} (\bibinfo{year}{2002}), \eprint{hep-ph/0210370}.

\bibitem[{\citenamefont{Obukhov et~al.}(1997)\citenamefont{Obukhov, Vlachynsky,
  Esser, and Hehl}}]{Obukhov:1997}
\bibinfo{author}{\bibfnamefont{Y.~N.} \bibnamefont{Obukhov}},
  \bibinfo{author}{\bibfnamefont{E.~J.} \bibnamefont{Vlachynsky}},
  \bibinfo{author}{\bibfnamefont{W.}~\bibnamefont{Esser}}, \bibnamefont{and}
  \bibinfo{author}{\bibfnamefont{F.~W.} \bibnamefont{Hehl}},
  \bibinfo{journal}{Phys. Rev.} \textbf{\bibinfo{volume}{D56}},
  \bibinfo{pages}{7769} (\bibinfo{year}{1997}).

\bibitem[{\citenamefont{Babourova and Frolov}(2003)}]{Babourova:2003}
\bibinfo{author}{\bibfnamefont{O.~V.} \bibnamefont{Babourova}}
  \bibnamefont{and} \bibinfo{author}{\bibfnamefont{B.~N.}
  \bibnamefont{Frolov}}, \bibinfo{journal}{Class. Quantum Grav.}
  \textbf{\bibinfo{volume}{20}}, \bibinfo{pages}{1423} (\bibinfo{year}{2003}),
  \eprint{gr-qc/0209077}.

\bibitem[{\citenamefont{Puetzfeld and Chen}(2004)}]{Puetzfeld:2004}
\bibinfo{author}{\bibfnamefont{D.}~\bibnamefont{Puetzfeld}} \bibnamefont{and}
  \bibinfo{author}{\bibfnamefont{X.-L.} \bibnamefont{Chen}},
  \bibinfo{journal}{Class. Quant. Grav.} \textbf{\bibinfo{volume}{21}},
  \bibinfo{pages}{2703} (\bibinfo{year}{2004}), \eprint{gr-qc/0402026}.

\bibitem[{\citenamefont{Randall and
  Sundrum}(1999{\natexlab{a}})}]{Randall:1999}
\bibinfo{author}{\bibfnamefont{L.}~\bibnamefont{Randall}} \bibnamefont{and}
  \bibinfo{author}{\bibfnamefont{R.}~\bibnamefont{Sundrum}},
  \bibinfo{journal}{Phys. Rev. Lett.} \textbf{\bibinfo{volume}{83}},
  \bibinfo{pages}{3370} (\bibinfo{year}{1999}{\natexlab{a}}),
  \eprint{hep-ph/9905221}.

\bibitem[{\citenamefont{Randall and
  Sundrum}(1999{\natexlab{b}})}]{Randall:1999b}
\bibinfo{author}{\bibfnamefont{L.}~\bibnamefont{Randall}} \bibnamefont{and}
  \bibinfo{author}{\bibfnamefont{R.}~\bibnamefont{Sundrum}},
  \bibinfo{journal}{Phys. Rev. Lett.} \textbf{\bibinfo{volume}{83}},
  \bibinfo{pages}{4690} (\bibinfo{year}{1999}{\natexlab{b}}),
  \eprint{hep-th/9906064}.

\bibitem[{\citenamefont{Godlowski and Szydlowski}(2004)}]{Godlowski:2004}
\bibinfo{author}{\bibfnamefont{W.}~\bibnamefont{Godlowski}} \bibnamefont{and}
  \bibinfo{author}{\bibfnamefont{M.}~\bibnamefont{Szydlowski}},
  \bibinfo{journal}{Gen. Rel. Grav.} \textbf{\bibinfo{volume}{36}},
  \bibinfo{pages}{767} (\bibinfo{year}{2004}), \eprint{astro-ph/0404299}.

\bibitem[{\citenamefont{Szydlowski and Krawiec}(2004)}]{Szydlowski:2004c}
\bibinfo{author}{\bibfnamefont{M.}~\bibnamefont{Szydlowski}} \bibnamefont{and}
  \bibinfo{author}{\bibfnamefont{A.}~\bibnamefont{Krawiec}},
  \bibinfo{journal}{Phys. Rev.} \textbf{\bibinfo{volume}{D70}},
  \bibinfo{pages}{043510} (\bibinfo{year}{2004}), \eprint{astro-ph/0305364}.

\bibitem[{\citenamefont{Kamionkowski et~al.}(1994)\citenamefont{Kamionkowski,
  Spergel, and Sugiyama}}]{Kamionkowski:1994}
\bibinfo{author}{\bibfnamefont{M.}~\bibnamefont{Kamionkowski}},
  \bibinfo{author}{\bibfnamefont{D.~N.} \bibnamefont{Spergel}},
  \bibnamefont{and} \bibinfo{author}{\bibfnamefont{N.}~\bibnamefont{Sugiyama}},
  \bibinfo{journal}{Astrophys. J.} \textbf{\bibinfo{volume}{426}},
  \bibinfo{pages}{L57} (\bibinfo{year}{1994}), \eprint{astro-ph/9401003}.

\bibitem[{\citenamefont{Riess et~al.}(2004)}]{Riess:2004}
\bibinfo{author}{\bibfnamefont{A.~G.} \bibnamefont{Riess}} \bibnamefont{et~al.}
  (\bibinfo{collaboration}{Supernova Search Team}),
  \bibinfo{journal}{Astrophys. J.} \textbf{\bibinfo{volume}{607}},
  \bibinfo{pages}{665} (\bibinfo{year}{2004}), \eprint{astro-ph/0402512}.

\bibitem[{\citenamefont{Peebles and Ratra}(2003)}]{Peebles:2003}
\bibinfo{author}{\bibfnamefont{P.~J.~E.} \bibnamefont{Peebles}}
  \bibnamefont{and} \bibinfo{author}{\bibfnamefont{B.}~\bibnamefont{Ratra}},
  \bibinfo{journal}{Rev. Mod. Phys.} \textbf{\bibinfo{volume}{75}},
  \bibinfo{pages}{559} (\bibinfo{year}{2003}), \eprint{astro-ph/0207347}.

\bibitem[{\citenamefont{Vishwakarma and Singh}(2003)}]{Vishwakarma:2003}
\bibinfo{author}{\bibfnamefont{R.~G.} \bibnamefont{Vishwakarma}}
  \bibnamefont{and} \bibinfo{author}{\bibfnamefont{P.}~\bibnamefont{Singh}},
  \bibinfo{journal}{Class. Quantum Grav.} \textbf{\bibinfo{volume}{20}},
  \bibinfo{pages}{2033} (\bibinfo{year}{2003}), \eprint{astro-ph/0211285}.

\bibitem[{\citenamefont{Spergel et~al.}(2003)}]{Spergel:2003}
\bibinfo{author}{\bibfnamefont{D.~N.} \bibnamefont{Spergel}}
  \bibnamefont{et~al.}, \bibinfo{journal}{Astrophys. J. Suppl.}
  \textbf{\bibinfo{volume}{148}}, \bibinfo{pages}{175} (\bibinfo{year}{2003}),
  \eprint{astro-ph/0302209}.

\bibitem[{\citenamefont{Page et~al.}(2003)}]{Page:2003}
\bibinfo{author}{\bibfnamefont{L.}~\bibnamefont{Page}} \bibnamefont{et~al.},
  \bibinfo{journal}{Astrophys. J. Suppl.} \textbf{\bibinfo{volume}{148}},
  \bibinfo{pages}{233} (\bibinfo{year}{2003}), \eprint{astro-ph/0302220}.

\bibitem[{\citenamefont{de~Bernardis et~al.}(2002)}]{deBernardis:2002}
\bibinfo{author}{\bibfnamefont{P.}~\bibnamefont{de~Bernardis}}
  \bibnamefont{et~al.}, \bibinfo{journal}{Astrophys. J.}
  \textbf{\bibinfo{volume}{564}}, \bibinfo{pages}{559} (\bibinfo{year}{2002}),
  \eprint{astro-ph/0105296}.

\bibitem[{\citenamefont{Chaboyer and Krauss}(2003)}]{Chaboyer:2003}
\bibinfo{author}{\bibfnamefont{B.}~\bibnamefont{Chaboyer}} \bibnamefont{and}
  \bibinfo{author}{\bibfnamefont{L.~M.} \bibnamefont{Krauss}},
  \bibinfo{journal}{Science} \textbf{\bibinfo{volume}{299}},
  \bibinfo{pages}{65} (\bibinfo{year}{2003}).

\bibitem[{\citenamefont{Puetzfeld}(2003)}]{Puetzfeld:2003}
\bibinfo{author}{\bibfnamefont{D.}~\bibnamefont{Puetzfeld}},
  \bibinfo{type}{Ph.D. diss.}, \bibinfo{school}{University of Cologne}
  (\bibinfo{year}{2003}).

\bibitem[{\citenamefont{Puetzfeld et~al.}(2005)\citenamefont{Puetzfeld,
  Pohl, and Zhu}}]{Puetzfeld:2005}
\bibinfo{author}{\bibfnamefont{D.}~\bibnamefont{Puetzfeld}},
  \bibinfo{author}{\bibfnamefont{M.} \bibnamefont{Pohl}},
  \bibnamefont{and} \bibinfo{author}{\bibfnamefont{Z.~H.}~\bibnamefont{Zhu}},
  \bibinfo{journal}{Astrophys. J.} \textbf{\bibinfo{volume}{619}},
  \bibinfo{pages}{657} (\bibinfo{year}{2005}), \eprint{astro-ph/0407204}.

\end{thebibliography}
\end{document}